# A Low-Temperature Atomic Layer Deposition Liftoff Method
## for Microelectronic and Nanoelectronic Applications

M. J. Biercuk[*], D. J. Monsma, C. M. Marcus,
*Department of Physics, Harvard University, Cambridge, MA 02138*

J. S. Becker, R. G. Gordon
*Department of Chemistry and Chemical Biology, Harvard University, Cambridge, MA 02138*

We report a novel method for depositing patterned dielectric layers with sub-micron features using atomic layer deposition (ALD). The patterned films are superior to sputtered or evaporated films in continuity, smoothness, conformality, and minimum feature size. Films were deposited at 100–150 C using several different precursors and patterned using either PMMA or photoresist. The low deposition temperature permits uniform film growth without significant outgassing or hardbaking of resist layers. A lift-off technique presented here gives sharp step edges with edge roughness as low as ~10 nm. We also measure dielectric constants ($\kappa$) and breakdown fields for the high-$\kappa$ materials aluminum oxide ($\kappa \sim 8$–9), hafnium oxide ($\kappa \sim 16$–19) and zirconium oxide ($\kappa \sim 20$–29), grown under similar low temperature conditions.

[*] To whom correspondence should be addressed. Electronic mail:
biercuk@physics.harvard.edu

A variety of applications require thin film oxides as gate dielectrics, simple insulators, or protective coatings. The push for dielectric layers exhibiting high conformality, uniform stoichiometry and thickness, large breakdown fields, and high dielectric constants has motivated a search for alternatives to $SiO_2$ and associated deposition techniques [1,2]. Chemical vapor deposition (CVD) provides highly uniform films on the wafer scale but requires high growth temperatures, which can damage underlying layers as well as polymer resists [3]. Sputtering and evaporation can be performed at lower temperatures, but often produce dielectric films that suffer from pinholes, poor conformality, and poor adhesion to the substrate [3].

An emerging deposition technique that offers relatively precise control of composition, conformality over high-aspect-ratio structures, and thickness control is atomic layer deposition (ALD) [4]. ALD is a self-limiting deposition process where separate precursor gases for a target material are sequentially dosed into a vacuum chamber under computer control: The first precursor chemisorbs to the substrate surface and the chamber is then purged. The second precursor is then introduced and reacts with the first, yielding a monolayer of film. The cycle then continues by reintroducing the first precursor, et cetera [5]. Substantial work has been invested to develop ALD processes that yield high quality films and use precursor gases that do not chemically damage preexisting device structures [5-7].

Previously, a significant shortcoming of ALD was that, like CVD, patterning of dielectrics required a *subtractive* process, in which whole layers were deposited, and patterning was done by etching. This limitation arose from the need to use deposition temperatures exceeding 300 C, which destroyed resist layers or caused them to outgas and disrupt film growth. In precise applications, etch steps are often unacceptable as they can damage other device structures. Moreover, it is difficult to pattern fine features by etching; first, because any subtractive process requires leaving a patch of resist on the substrate the same size as the desired feature, and second, because low energy dielectric etches (which do not damage the device) are generally isotropic, causing feature shape change and undercutting [3]. In contrast, liftoff patterning allows one to cut a "slit" in a resist layer and deposit material only where needed. It is therefore desirable to develop liftoff processes for dielectrics similar to those used for metallization. High quality dielectric films patterned by liftoff would be of great value not only in the semiconductor industry, but also in optical applications [8,9], as catalysts [10,11], and as protective coatings [12,13].

In this Letter we demonstrate a process that allows ALD-grown dielectric films to be patterned using liftoff. Examples described in detail are the high–$\kappa$ materials aluminum oxide ($Al_2O_3$) and hafnium oxide ($HfO_2$). The ALD process employed operates at low temperature and uses non-corrosive precursor gases. The patterned films are uniform in thickness with deviations ~1 nm, and are conformal to underlying device structures. ALD liftoff is demonstrated for both photolithography and electron-beam lithography, yielding films with patterned features below 100 nm. We have also measured the dielectric constants and breakdown fields of comparably grown unpatterned films of $Al_2O_3$, $HfO_2$ and $ZrO_2$, finding $\kappa \sim 8.2 - 9$ for $Al_2O_3$, $\kappa \sim 16.3 - 18.5$ for $HfO_2$, and $\kappa \sim 20 - 29$ for $ZrO_2$, at various film thicknesses and measurement temperatures. All films measured exhibit breakdown fields between 5.6 and 9.5 MV/cm, varying with material, film thickness and temperature (see Table 1).

The photolithographic ALD process consisted of the following steps. First, cm-scale pieces of a polished Si wafer with 1μm of thermally grown oxide were cleaved, cleaned (5 minutes in each of tricholoroethlyene, acetone, methanol) and baked for 5 minutes at 160 C to drive off solvent residues. Next, Shipley 1813 or 1818 photoresist was spun onto the samples, after which they were baked for 2 minutes at 120 C and exposed through a photomask with large features (>10μm). Patterns were developed using Microposit CD-26 Metal-Ion-Free developer (tetramethyl ammonium hydroxide) and cleaned for 30 s in 100 W oxygen plasma at 700 mtorr. Thin films were then grown on these samples via the novel ALD process, as described below. The electron-beam ALD process, used for fabricating fine features, began with similar Si samples, cleaved and cleaned using the same three-solvent rinse followed by a 2 minute bake at 180 C. A bilayer of 200k PMMA and 950k PMMA was spun onto a sample and baked for 15 min at 180 C for each layer, yielding a total PMMA thickness ~ 350 nm. Fine-line patterns were written and developed in a solution of isopropanol (75%) methyl isobutyl ketone (24%), and methyl ethyl ketone (1%).

The ALD procedure used for both the photolithographic and electron-beam liftoff processes employed highly reactive metal amide precursors (tetrakis(dimethylamido)-hafnium (IV) and $H_2O$ for $HfO_2$; tetrakis(dimethylamido)zirconium (IV) and $H_2O$ for $ZrO_2$ [5-7, 14-16], and trimethylaluminum and $H_2O$ for $Al_2O_3$ [17-19]). Samples were placed in a stainless steel tube furnace and heated to 100–150 C. The cycle of precursors was then started, with nitrogen purges between each step. In order to achieve low-temperature deposition with uniform thickness, the nitrogen purge time needed to be lengthened (from ~5 s, used for the 300 C process, to ~120 s) to prevent physisorption and to remove unreacted gas-phase precursors. Film thicknesses ranged from 2.5 to 100 nm.

Despite the reduced temperature and lengthened total deposition time, the films appear similar in composition to those grown at higher temperatures, although some important differences exist. First, while surface roughness of these films is typically ~5% of total film thickness for high temperature deposition (>200 C) it is less than 1% total film thickness for deposition temperatures below 150 C, except where limited by substrate roughness [20]. Second, X-ray diffraction data indicate that 100 nm thick films of unpatterned $HfO_2$ grown at or below 100 C are completely amorphous while those grown at higher temperatures show some crystallinity (<10% for growth temperature up to 200 C) [20]. Low-temperature grown $ZrO_2$ films characterized in the same manner show an increase in crystallinity from 10% to 60% as the deposition temperature increases from 100 to 150 C [20]. It is worth noting that amorphous dielectric films are desirable for applications as gate dielectrics due to their smoothness and high breakdown fields compared to polycrystalline films [21,22].

Following the growth step, the liftoff procedure was carried out by immersing samples in acetone for times ranging from 10 m to 2 h. To allow the acetone to penetrate the conformal dielectric layer and attack the resist below, it was necessary to manually scratch the surface of the film. While still immersed in acetone, ~1 s pulses from an ultrasonic bath were used to dislodge remaining sections of resist.

Atomic force microscope images (Fig. 1) show that the resulting patterned films have surface roughness comparable to that of the $SiO_2$ substrate (~1 nm), and sharp step edges. Deviation of the edge from a straight line is limited by the photolithography and not film deposition or liftoff. This was verified by examining metal lines deposited in similar patterns, as shown in Fig. 4. Micrographs of patterned ALD films on $SiO_2$ (Figs. 1(a) and 2(a)) show edge roughness ~10 nm for electron-beam patterning and ~100 nm edge roughness for photolithographic patterning (Figs. 1(b) and 2(b)).

Having established liftoff of ALD we next tested whether submicron features could be patterned with this method. Figure 3 shows an SEM image of a device geometry featuring lines of dielectric patterned via electron-beam lithography, with smallest dimensions below 100 nm. We have also fabricated complicated multilayer device geometries in which metallic layers are partially coated with patterned ALD films, followed by patterned metallic overlayers. SEM analysis shows (Fig. 4) that patterned ALD films running over metallic lines are highly conformal around the metal line edge and at the metal-substrate interface.

The dielectric constants and breakdown voltages of unpatterned dielectic films grown by low-temperature ALD as described above were measured as follows. Films of $Al_2O_3$, $HfO_2$ and $ZrO_2$ were grown on Si-SiO$_2$ substrates with 20 nm Ti + 50 nm Pt electrodes deposited by electron-beam evaporation. The Ti-Pt formed films with surface roughness comparable to that of the substrate. In addition, the large work function of Pt created a high electron barrier height. ALD films were deposited at 150 C and subsequently 50 nm Pt was evaporated through a shadow mask to form a top electrode of dimension ~ 200 μm × 200 μm. We find that these film show good adhesion to the Pt underlayer, characteristic of most ALD processes, and unlike sputtered or evaporated dielectric films. These tri-layer structures formed parallel-plate capacitors, which were characterized in a vacuum probe station at 20 K and room temperature. A 1 kΩ resistor was placed in series with these test devices, and digital lock-ins were used to measure the voltage drops across both the resistor and the test device. The circuit was voltage biased using a function generator with an excitation of ~100 mV at 1 kHz. Voltages across the resistor ($V_R$) and test device ($V_C$) were used to measure the capacitance of the test device, $C = V_R \left( 2\pi f R V_C \right)^{-1}$ and hence the dielectric constant of the film, $\kappa = Cd / \left( \varepsilon_0 A \right)$ (A is the device area; d is the film thickness). Dielectric constants $\kappa \sim 20$–$29$ are found for $ZrO_2$, $\kappa \sim 16$–$19$ for $HfO_2$, and $\kappa \sim 7$–$8$ for $Al_2O_3$ (see Table 1). Breakdown fields $E_{BD} = V_{BD}/d$ were found by applying an increasing dc bias until the onset of a large leakage current was observed at $V_{BD}$. Values obtained are in the range $E_{BD} \sim 6 - 9$ MV/cm for all three materials, approaching the breakdown fields for high-quality $SiO_2$ films. Resulting values for dielectric constants, breakdown fields, and induced charge densities as calculated from the measured breakdown fields are presented in Table 1 for varying thicknesses and measurement temperatures of the materials. It is interesting to note that the values we obtain for breakdown fields in these devices are two to three times higher than those previously reported in the literature for $HfO_2$ and $ZrO_2$ [23-26]. We believe the difference is due to the low-temperature growth process, which produces amorphous films.

In summary, we have developed a method for patterning dielectric films using a liftoff process with low-temperature atomic layer deposition (ALD). With this technique, we have fabricated dielectric layers with features down to 80 nm. We have also measured the dielectric constants and breakdown voltages of several high–$\kappa$ dielectric materials fabricated using this low-temperature ALD process. The film quality and achievable pattern dimensions make this process particularly suitable for microelectronic and nanoscale applications.

This work was supported by funding from the National Science Foundation through the Harvard MRSEC, NSF-DMR-0213805 and NSF-CTS 0236584, and the Army Research Office, under DAAD19-02-1-0039 and DAAD19-02-1-0191. MJB acknowledges support from an NSF Graduate Research Fellowship.

**Figure Captions:**

**Fig. 1**. Atomic force micrographs of patterned ALD oxides on a Si/SiO$_2$ substrate. a) A 15 nm thick narrow line of oxide patterned via electron-beam lithography necks down from 5$\mu$m to 1$\mu$m (top to bottom). Note the vertical edge profile on both sides of the line and edge smoothness. b) AFM surface plot of a ~22 nm thick mesa of ALD oxide patterned by photolithography shows a well defined and highly vertical step. The waviness seen in the edge is limited by the photolithography.

**Fig. 2**. Scanning electron micrograph showing the smooth edge profiles of ALD patterned via a) electron-beam lithography, b) photolithography. Surface roughness was ~1 nm as analyzed by AFM (Shadows result from high-angle imaging).

**Fig. 3**. SEM image of 15 nm thick HfO$_2$ on Si/SiO$_2$, patterned by electron beam lithography. Device critical dimensions ~80 nm as measured using the SEM. Inset: region of the device showing smallest features.

**Fig. 4**. Multilayer structure (30 nm ALD oxide and 70 nm Ti/Au, both patterned using photolithographic liftoff) showing high conformality of the ALD around the edge of the Ti/Au.

**Table 1**. Properties of several high-$\kappa$ materials grown using the same low-temperature ALD process as used for liftoff, measured at 20K and room temperature: breakdown field, $E_{BD} = V_{BD}/d$  ($V_{BD}$ is breakdown voltage, d is film thickness), dielectric constant $\kappa$ (see text), and charge density at breakdown, $Q_{BD} = CV_{BD}$.

**References:**

[1] G.D. Wilk, R.M. Wallace, J.M. Anthony, J. Appl. Phys **89**, 5243 (2001).

[2] M. Gutsche, H. Seidl, J. Leutzen, A. Birner, T. Hecht, S. Jakschik, M. Kerber, M. Leonhardt, P. Moll, T. Pompl, H. Reisinger, S. Rongen, A. Saenger, U Schroeder, B. Sell, A. Wahl, D. Schumann, Technical Digest of the Internatinal Electronic Device Meeting, 18.6.1, IEEE Piscataway, NJ (2001).

[3] M. Quirk and J. Serda, Semiconductor Manufacturing Technology, Prentice-Hall, NJ (2001).

[4] M. Leskela, M. Ritala, J. Phys . IV Fr. **9-Pr8**, 837 (1999).

[5] D. M. Hausmann, E. Kim, J. S. Becker, R. G. Gordon, Chem. Mater. **14**, 4350 (2002).

[6] J. S. Becker, Thesis, Dept of Chemistry, Harvard University, (2002).

[7] D.M. Hausmann, Thesis, Dept of Chemistry, Harvard University, (2002).

[8] A.J. Waldorf, J.A. Dobrowolski, B.T. Sullican, L.M. Plante, Appl. Opt. **32**, 5583 (1993).

[9] C. Urlacher, J. Mugnier, J. Raman, Spec. **27**, 785 (1996).

[10] J. Nawrocki, M.P. Rigney, A. McGormick, P.W. Carr, J. Chromatogr. **A657**, 229 (1988).

[11] K. Asakura, M. Aoki, Y. Iwasawa, Catal. Lett. **1**, 394 (1988).

[12] H. Ibegazene, S. Aplerine, C. Diot, J. Mater. Sci. **30**, 938 (1995).

[13] J. Wang, H. Li, R. Stevens, J. Mater. Sci. **27**, 5397 (1992).

[14] R. G. Gordon, D. Hausmann, E. Kim, J.  Shepard, Chemical Vapor Deposition **9**, 73 (2003).

[15] R. G. Gordon, J. Becker, D. Hausmann, S. Suh, Materials Res. Soc. Symp. Proc. **670** (Gate Stack and Silicide Issues in Silicon Processing II), K2.4/1-K2.4/6 (2002).

[16] R. G. Gordon, J. Becker, D. Hausmann, S. Suh, Chem. Mater. **13**, 2463 (2001).

[17] A. W. Ott, J. W. Klaus, J. M Johnson, and. S. M. George, Thin Solid Films **292**, 135 (1997).

[18] R. Matero, A. Rahtu, M. Ritala, M. Leskelä, and. T. Sajavaara, Thin Solid Films **368**, 1 (2000).

[19] **A.** Rahtu, T. Alaranta, M Ritala,  Langmuir **17**, 6506 (2001).

[20] D.M. Hausmann, R.G. Gordon, J. Crystal Growth **249**, 251 (2003).

[21] P. Balk, J. Non-Cryst. Sol. **187**,1 (1995).

[22] K. Kukli, J. Ihanus, M. Ritala, M. Leskela, J. Electro-chem. Soc. **144**, 300 (1997).

[23] J. McPherson, J. Kim, A. Shanware, H. Mogul, J. Rodriguez, IEDM Technical Digest, **XX** (2002).

[24] T. Ma, Solid-State IC Technology Conference Proceedings **297** (2001).

[25] L. Kang, et.al., IEEE Electron Device Letters **21**, 181 (2000).

[26] L. Manchanda, et.al., IEDM Technical Digest **23** (2000).

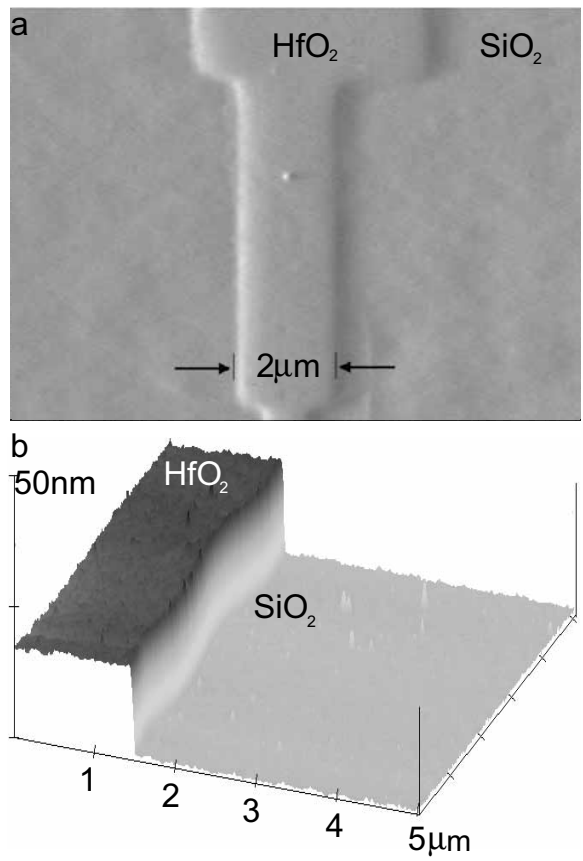

Fig. 1

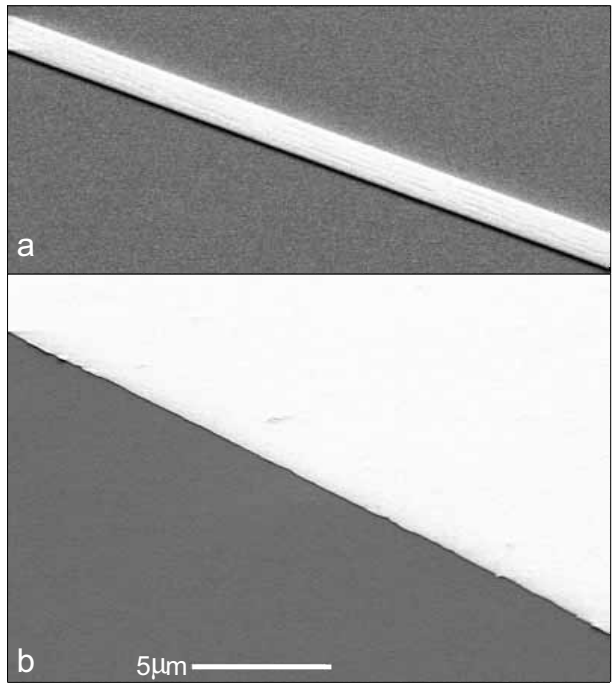

a

b

5μm

Fig. 2

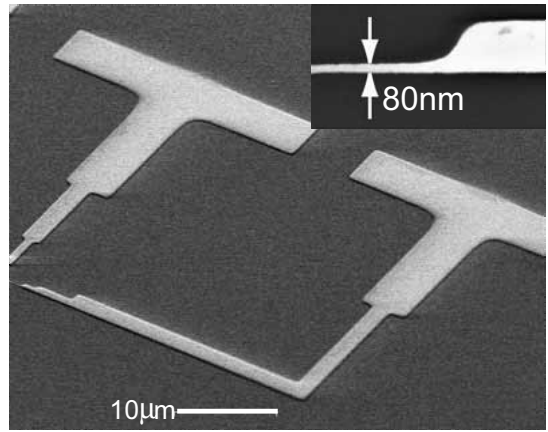

80nm

10µm

Fig. 3

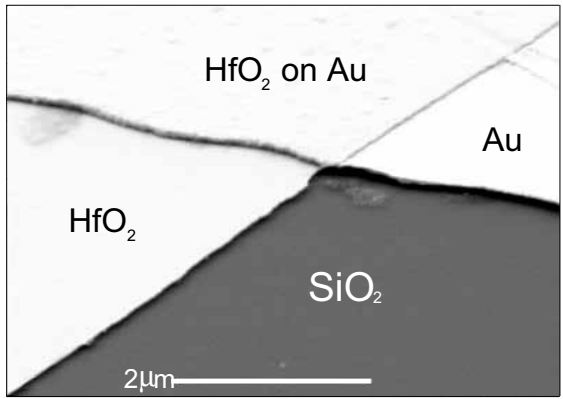

Fig. 4

| Material | Thickness | Temperature | $E_{BD}$ | $\kappa$ | $Q_{BD}$ |
|---|---|---|---|---|---|
| $Al_2O_3$ | 2.5 nm | RT | 8 MV/cm | 9 | 6.4 $\mu C/cm^2$ |
| $Al_2O_3$ | 10 | RT | 8.3 | 8.8 | 6.5 |
| $Al_2O_3$ | 25 | RT | 8.2 | 8.2 | 6.0 |
| $Al_2O_3$ | 50 | RT | 7.6 | 8.9 | 6.0 |
| $ZrO_2$ | 25 | RT | 5.6 | 20 | 9.9 |
| $ZrO_2$ | 100 | RT | 6 | 29 | 15.5 |
| $ZrO_2$ | 50 | 20K | 8.2 | 29 | 21 |
| $ZrO_2$ | 100 | 20K | 9.5 | 26 | 22 |
| $HfO_2$ | 10 | RT | 6.5 | 17 | 9.7 |
| $HfO_2$ | 25 | RT | 7.4 | 18.5 | 12 |
| $HfO_2$ | 25 | 20K | 8.4 | 16.3 | 12.1 |

Table 1.